# Pathfinder: Exploring Path Diversity for Assessing Internet Censorship Inconsistency

Xiaoqin Liang
Old Dominion University
Norfolk, Virginia, USA
xlian001@odu.edu

Guannan Liu
Colorado School of Mines
Golden, Colorado, USA
guannan.liu@mines.edu

Lin Jin
University of Delaware
Newark, Delaware, USA
linjin@udel.edu

Shuai Hao
Old Dominion University
Norfolk, Virginia, USA
shao@odu.edu

Haining Wang
Virginia Tech
Arlington, Virginia, USA
hnw@vt.edu

## ABSTRACT

Internet censorship is typically enforced by authorities to achieve information control for a certain group of Internet users. So far existing censorship studies have primarily focused on country-level characterization because (1) in many cases, censorship is enabled by governments with nationwide policies and (2) it is usually hard to control how the probing packets are routed to trigger censorship in different networks inside a country. However, the deployment and implementation of censorship could be highly diverse at the ISP level. In this paper, we investigate Internet censorship from a different perspective by scrutinizing the diverse censorship deployment inside a country. Specifically, by leveraging an end-to-end measurement framework, we deploy multiple geo-distributed back-end control servers to explore various paths from one single vantage point. The generated traffic with the same domain but different control servers' IPs could be forced to traverse different transit networks, thereby being examined by different censorship devices if present. Through our large-scale experiments and in-depth investigation, we reveal that the diversity of Internet censorship caused by different routing paths inside a country is prevalent, implying that (1) the implementations of centralized censorship are commonly incomplete or flawed and (2) decentralized censorship is also common. Moreover, we identify that different hosting platforms also result in inconsistent censorship activities due to different peering relationships with the ISPs in a country. Finally, we present extensive case studies in detail to illustrate the configurations that lead to censorship inconsistency and explore the causes.

## 1 INTRODUCTION

Internet censorship has become increasingly pervasive as more governments and authorities rely on it to restrict users from accessing undesired content. In many cases, censorship policies are typically enforced by nations' authorities, resulting in that nationwide censorship could be largely consistent. Thus, most censorship studies aggregate the observed activities at the country level [29, 30, 40, 43, 44, 48, 50, 55]. However, the severity of censorship may be highly diverse at the ISP level, due to the difference of deployments and implementations, which makes the country-level characterization too coarse to draw reasonable conclusions for examining censorship deployment. Limited by the measurement methods, such diversity is mostly underestimated in previous work as researchers have no control on which transit networks or gateways the experiment packets will reach and thus cannot attribute inconsistent censorship behaviors inside a country to diverse censorship deployment.

Although some previous studies reported decentralized information control in certain specific countries [22, 51, 62], little research has examined the inconsistency of censorship enforcement in a systematic manner and at a global scale. One of the closest studies is [14], which leverages BGP churn to identify the Autonomous Systems (ASes) responsible for inducing censorship activities. However, it entirely relies on the measurement data from ICLab [40] that uses VPNs only, so it has limited coverage and may miss some key observations [30]. Moreover, the scale of network-level path diversity caused by BGP churn is limited and the occurrence of a path churn is random and cannot be controlled by the experiments. BreadCrumb [9] studies the DNS censorship changes due to router load balancing based on different packet parameters, *i.e.*, ephemeral source port and local bits of the source IP address. However, it is dedicated to investigating the impact of DNS censorship, and its routing changes relying on router load balancing only explore a small-scale routing variation inside a network.

In this paper, we perform an in-depth investigation on Internet censorship from a different perspective that is largely overlooked in previous studies, by which we scrutinize the inconsistency of censorship deployments and implementations inside a country. Specifically, we design and deploy a measurement framework called *Pathfinder* to identify inconsistent censorship activities experienced on different network paths from vantage points inside a country. Pathfinder leverages multiple geo-distributed back-end control servers as different destinations of probing packets to explore potential diverse routing paths. As such, the probing packets issued from the vantage points in the same country could be routed to different transit networks when using different destination IPs of control servers, resulting in diverse censorship behaviors due to inconsistent policies on different networks. In addition, our system design is also inspired by Disguiser [30], which achieves accurate censorship detection by providing the control server a static payload as the ground truth of server responses. We set up Pathfinder's control servers in the same way to accurately identify censorship activities without manual inspection.

Through Pathfinder, we perform a large-scale measurement study to understand the censorship inconsistency caused by diverse routing paths. We conduct a 60-week experiment (from Summer 2022 to Fall 2023) by acquiring more than 144K vantage points from 120 countries with detected censorship activities, and reveal that such a phenomenon is quite prevalent, where 91.7% of countries (110) experience various extents of inconsistency though many of them are commonly considered as having centralized censorship control. Especially, we observe that some paths can have a lower percentage of censorship activities than others, indicating that potential censorship circumvention could be explored. For instance, with the vantage points from India, the packets routed to the paths toward the control servers deployed in the Middle East experience much less censorship than other paths (8% vs. 40-60%).

Moreover, we uncover that different cloud hosting platforms also contribute to inconsistent censorship behaviors, due to their various peering connections or preferences, which could cause the packets to be routed to different transit networks that have different censorship deployments. On the other hand, direct peering between cloud platforms and the eyeball networks inside a country could allow certain censorship circumvention, because the packets would enter cloud providers' private networks before reaching the censorship devices deployed at the upstream networks. To illustrate an in-depth investigation for such cases, we perform extensive case studies for South Korea and India by examining their detailed censorship deployments.

In summary, this work makes the following contributions:

- Developing Pathfinder, a framework to simultaneously explore multiple potential routing paths for identifying censorship activities when probing packets from one node are routed to different ISPs or transit networks.
- Conducting extensive measurements and uncovering that the censorship deployments inside a country are largely inconsistent, even for many countries that are considered to have centralized censorship controls. We further show that certain paths could experience far less censorship, which can be exploited for potential censorship circumvention.
- Discovering that large cloud platforms can affect the occurrence of censorship due to peering connections. We leverage application traceroutes to perform case studies in two countries (South Korea and India) to collect detailed network paths and examine different peering configurations that lead to inconsistent censorship.

# 2 BACKGROUND

## 2.1 Censorship Techniques

Internet censorship can be achieved by various techniques [23], such as IP-layer or application-layer censorship. IP-based censorship examines the destination IP of a packet, and thus it could be easily evaded by the changes of service IP addresses. Besides, with the wide adoption of cloud and CDNs, the IP resources become increasingly shared and dynamic so that the IP-based blocking often results in collateral damage for legitimate services [20, 63]. Thus, application-layer censorship that can accurately block undesired content has been attracting more attention from both censorship regimes and research community, and is the focus of this work.

**Application-layer Censorship.** Application-layer censorship involves the information inspection inside the application-layer protocols, *e.g.*, domain-name-based blocking in DNS and HTTP(S), and keyword-based filtering in packet streams.

Domain names explicitly reveal the Internet resources a client intends to access, which allows a censorship device to prevent users from accessing online content that is prohibited by authorities. When a user device resolves a domain name to its web server's IP address through DNS, the censor can directly learn the accessed domain as DNS traffic is typically unencrypted. In addition, the HTTP Host header presents the domain that a client is visiting. Since the HTTP protocol is unencrypted, the censors can know exactly the requested domain. HTTPS encrypts all the HTTP packets after a TLS handshake so that the Host header is no longer visible to the censors, but commonly an SNI (Server Name Indication) extension is sent in plaintext in the Client Hello message to indicate which domain the client intends to visit.

In order to achieve finer control to prohibit undesired content, keyword-based censorship searches the sensitive keywords in unencrypted packet streams and disrupts the traffic when those predefined, forbidden keywords are detected. To do so, the censorship device intercepts and parses all HTTP traffic, and searches the keywords in certain locations of HTTP requests or responses such as request line, headers, or payload body [58].

**Interference Techniques.** The mechanisms used for blocking or filtering undesired Internet content also vary. To perform DNS manipulation for denying a domain's access, the censorship devices usually inject forged DNS responses that redirect user's requests to a censor-controlled address (*e.g.*, that shows a *blockpage* indicating the domain being prohibited), a non-routable private IP address, or a public IP address that will be prohibited by IP-layer blocking [25]. Additionally, in the case of TCP-based DNS, a censor typically injects an RST/FIN packet to tear down the TCP connection [30].

In the HTTP/HTTPS blocking, when a prohibited domain is detected (*i.e.*, from the HOST header in unencrypted HTTP messages or the plaintext Client Hello message in the initial HTTPS handshake), the censor can simply drop the request, causing a timeout on the client side. Alternatively, the censors could also tear down the connection by sending RST/FIN packets to both the client and the server of the requested domain, or respond to the client with a dedicated blockage that indicates the domain blocking policy.

**Inbound and Outbound Censorship.** Censorship devices can be deployed to examine different directional traffic, *i.e.*, inbound and outbound censorship [55]. Inbound censorship refers to that censorship devices monitor and intercept the traffic in the inbound direction where the traffic originates from the networks outside the country/region. On the contrary, in outbound censorship, censorship devices inspect the outbound traffic that originates internally and traverses toward the destination outside their networks. Compared to blocking undesired content detected inbound to the network, outbound censorship is more common as censorship policies typically aim to control how sensitive content can be accessed by the users within the censoring regions.

## 2.2 Application Traceroute

To better understand censorship, application traceroute has been explored to pinpoint the location of censorship devices and examine their behaviors [30, 50]. In doing so, application traceroute increments the TTL field in the probing packets, while the application payload is set to trigger the censorship. Before reaching the censor, the probing packets will be dropped and an ICMP Time Exceeded message will be returned. As the TTL increases and the probing packet reaches the censorship device, the sign of interference (*e.g.*, an RST/FIN packet) indicates the exact hop of censors.

# 3 PATHFINDER

In this section, we introduce our methodology and framework, Pathfinder, to systematically investigate censorship inconsistency by measuring censorship activities from diverse paths simultaneously. We first discuss the phenomenon of censorship inconsistency and the challenges of examining this problem through existing censorship measurement platforms. We then describe the design of Pathfinder, as well as special design considerations that eliminate potential noise data when conducting large-scale experiments.

## 3.1 The Censorship Inconsistency Problem

Understanding censorship activities often involves sending a probing request that could trigger the censorship from a vantage point inside a country or region and detecting network interference by analyzing abnormal responses. With such a setup, however, the detection of censorship entirely relies on responses observed on the vantage points' side, and thus no path information can be learned and no location of censorship devices can be inferred. Consequently, censorship behaviors are usually characterized by aggregating the results at the country level. However, the censorship policies are usually deployed at the ISP level and different ISPs may have different censorship implementations, making such characterization inaccurate to unveil inconsistent censorship behaviors in a country.

Furthermore, traceroute or application-layer traceroute has been utilized to examine network path and censorship devices [30, 50], but it would suffer from limited coverage of measured networks for identifying inconsistent censorship behaviors. Figure 1 shows how such inconsistency is typically overlooked by existing measurement studies. To trigger censorship devices situated on the network path, the probing packets, including traceroutes, are sent through the vantage points in a country to the test domain. However, as the probing packets are all toward the same destination, the packets issued from distributed vantage points may converge to a certain few upstream transit networks, *e.g.*, AS 1 in Figure 1. On the other hand, AS 2 may implement different censorship policies for the test domain, but the regular measurements cannot identify such behavior due to routing configurations.

## 3.2 System Design

To systematically examine the censorship inconsistency, we design and deploy a measurement framework, Pathfinder, to detect censorship in various paths. Our design rationale is that if we are able to direct probing packets to traverse different networks inside a country, we can understand the censorship deployments from a different perspective in a finer-grained manner by characterizing the impact of diverse routing paths on the changes of censorship behaviors.

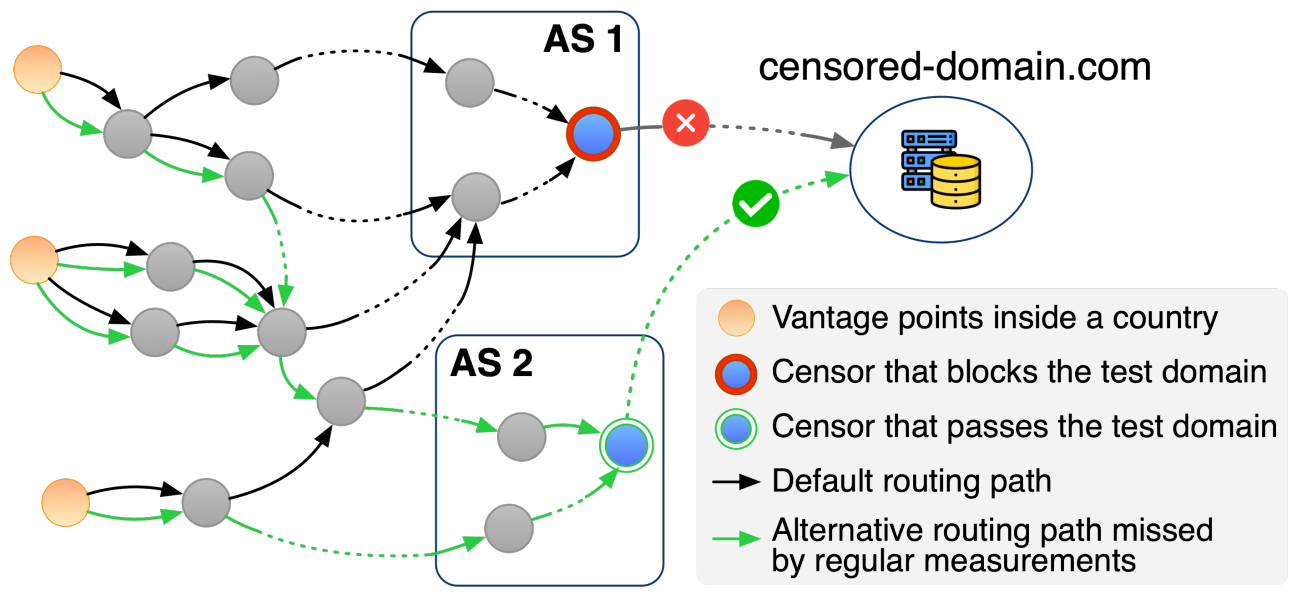

**Figure 1: Inconsistent censorship in different paths due to different censorship implementations.**

**Overview.** Figure 2 illustrates the system design of Pathfinder. At a higher level, Pathfinder consists of a set of vantage points distributed across the world and a set of back-end control servers that serve as the destination of probing packets. Also, a client feeds a country-specific test domain list into the vantage points to construct HTTP requests and schedule the measurements for initiating HTTP connections with each different control server, in which test domain is embedded into the HTTP packet header field to trigger censorship. On the other hand, control servers respond to all received HTTP requests with a static payload. Pathfinder then collects the response and connection status for each probing test at both client and control server sides for analysis.

**Vantage Points.** In order to issue probing requests and conduct experiments, we need to acquire a set of vantage points (VPs) distributed across the world. Similar to prior study [30], we leverage SOCKS IP proxies to issue probing requests. SOCKS IP services enable customers to proxy traffic through an entry node called the gateway server in the service providers' infrastructure. The gateway server then further forwards the traffic to one of exit nodes, which eventually relays traffic to the destination as the traffic source (*i.e.*, the vantage points in the Pathfinder's framework).

There are multiple types of SOCKS proxies available such as residential IP proxies, ISP IP proxies, and datacenter IP proxies. We performed extensive tests on these different services from different platforms to validate their usability. We confirmed that residential networks have been experiencing more aggressive censorship than the vantage points in commercial infrastructures such as data centers [40, 61], and thus vantage points from residential IP proxies (RESIP) can provide a more comprehensive and representative coverage for examining censorship deployments. To this end, we eventually choose and subscribe to Proxyrack [45], a popular and stable RESIP service that has been extensively analyzed and used in previous studies [29, 30, 37]. In addition, we extensively discuss the ethical considerations of using RESIP and our strategies to mitigate ethical concerns in Appendix § 3.4.

**Identified Censored Domain List.** As Pathfinder's main goal is to examine the censorship inconsistency, we use the existing list of identified censored domains to reduce the measurement efforts that purely aim to detect censorship occurrence. We acquire such a list from Disguiser [30], which compiles a test domain list that consists of popular domains and sensitive domains extracted from Alexa's top 1,000 domains [4] and the Citizen Lab [15], and produces a country-specific list of censored domains [52] validated with Disguiser's measurements. By applying this censored-domain list to Pathfinder, in this study we only need to issue probing packets with those domains that have been observed to be censored by at least one censorship device in the country.

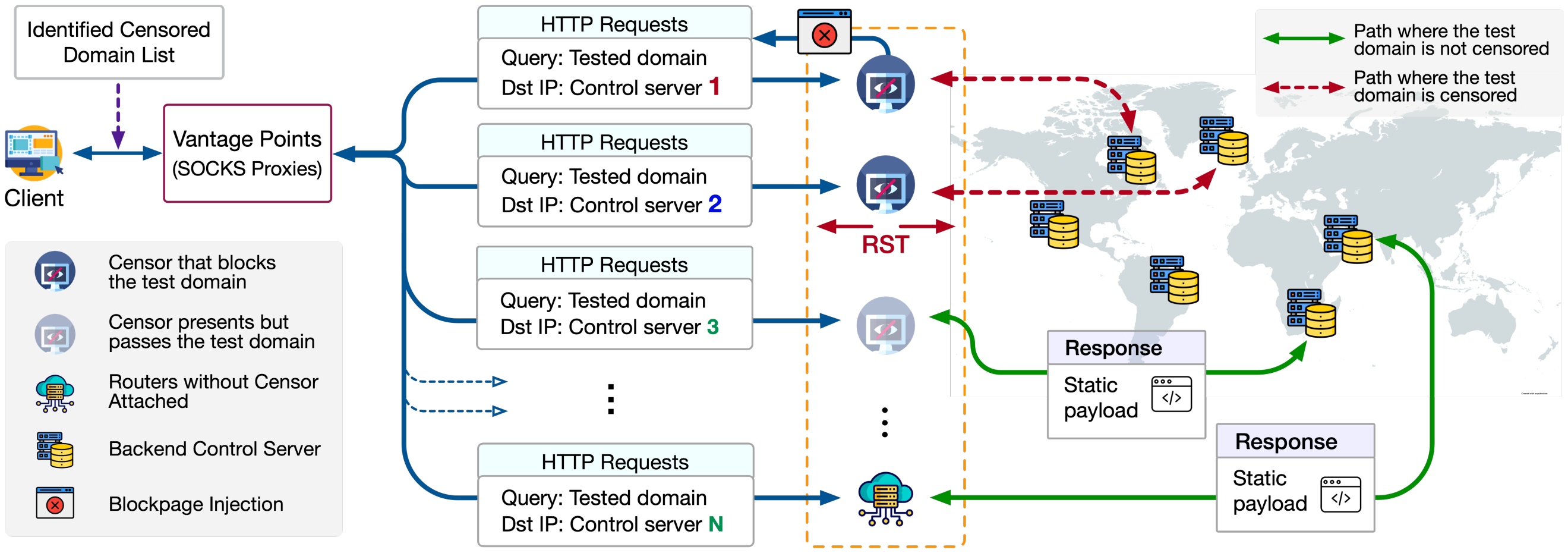

**Figure 2: System design of Pathfinder and its exploitation of censorship inconsistency in diverse paths.**

**HTTP Requests.** With each vantage point acquired from the RESIP, Pathfinder constructs a set of HTTP requests and sends them to trigger potential censorship on the network paths. In this study, we focus on the censorship for HTTP requests as it is the most prevalent censorship deployed by more countries [30, 40]. Each request is constructed as that the HOST header contains a domain name from the identified censored-domain list and the destination is set to one of the IP addresses of our control servers. We set a 5-second timeout for each probing request so that we do not need to wait an extended time if a censorship device drops our requests. Although prior study adopted a 15-second timeout [30], in our study, we observed that if we do not receive the response in 5 seconds, then it is highly unlikely to receive the response with a longer waiting time. Also, Pathfinder automatically retries four more times for the timeout requests before declaring it to be blocked by censorship, considering that some requests may be dropped due to network congestion.

**Control Server.** Control servers play a key role in identifying censorship activities while establishing path diversity for various HTTP requests. First, similar to [30], the control servers provide a static payload for each HTTP request received. Such static payload is unique in a way that it should not collide with other legitimate domain pages or block pages, thereby providing a ground truth for efficiently detecting the occurrence of censorship. In addition, Pathfinder's design establishes multiple control servers and each server is deployed at different locations, so that the HTTP requests connecting different control servers could be routed differently due to different destinations, as shown in Figure 2. We detail the control server setup for our experiments in Section 4.

## 3.3 Special Design Considerations

**Eliminating Cache Proxies.** To accurately identify censorship occurrence, we also need to consider a special scenario when a cache proxy is present on the network path. In this case, the cache proxy may directly respond to a probing request with the cached result for a test domain from previous connections, and the vantage point will receive a response rather than the valid static payload of control servers.

To eliminate such potential impact on our measurements, similar to [30], we perform a real-time cache proxy test for each vantage point before we include it in the experiments, removing those that potentially have cache proxies situated on the path. In doing so, we first set up two reference servers serving the same domain under our control but with distinct destinations and payloads of landing pages. Then, we instruct the vantage points to sequentially probe the two reference servers by fetching the landing pages. If the cache proxy is present, the second probing could receive a response associated with the page of the first reference server which is cached through the first probing test.

By this means, we can successfully exclude most vantage points that are possibly impacted by cache proxies. However, we find that there is still a small portion of vantage points receiving responses from potential intermediate caches. This could be because a cache server selectively caches the content it passes through or it may intercept the connections by independently performing DNS resolution so as to obtain a different IP address other than our control server's. Therefore, we perform an offline check to exclude those vantage points. We fetch the legitimate landing pages of the test domains and simply compare their title tags with the responses received by vantage points. If a vantage point has obtained a valid title tag from the legitimate page of a test domain, it implies that the page could be retrieved and returned by the intermediate cache proxies. In total, we exclude 164 acquired vantage points from our measurements by offline check.

**Eliminating Inbound Censors.** As mentioned in Section 2.1, censorship devices can appear as inbound censors or outbound censors.

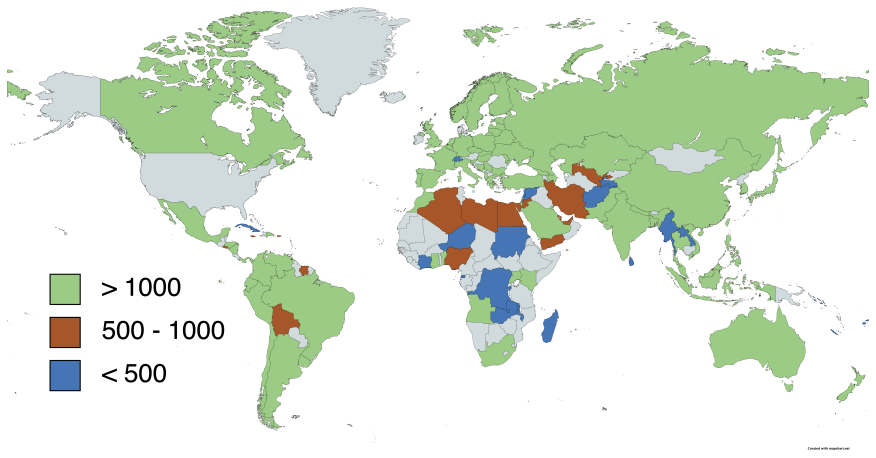

**Figure 3: Distribution of vantage points across all countries in §4.1.**

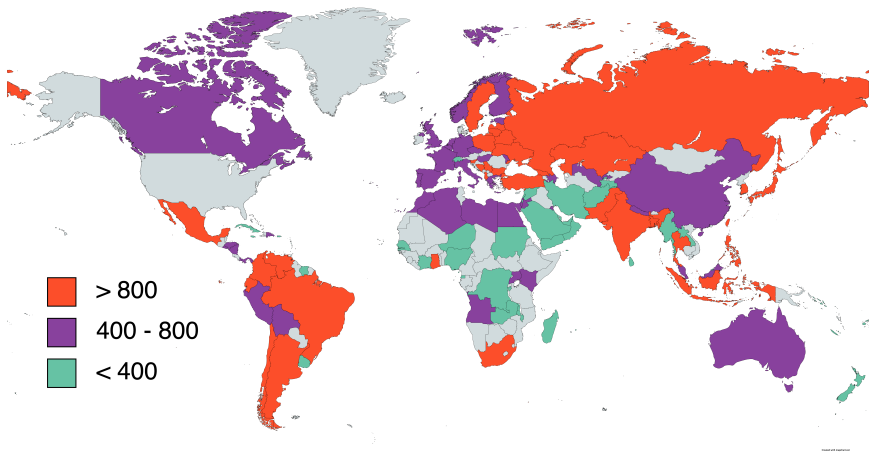

**Figure 4: Distribution of vantage points across all countries in § 4.2.**

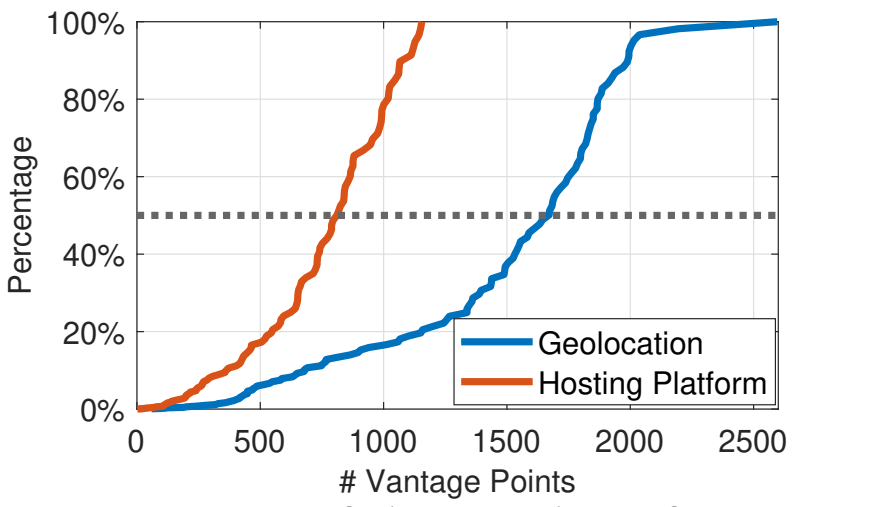

**Figure 5: CDF of the number of vantage points in § 4.1 and § 4.2.**

Since we mainly focus on the censorship enforced by the countries in which the vantage points located, the inbound censorship deployed by the countries where control servers are deployed may also be identified as censorship cases by Pathfinder, which could introduce false cases in our results. To avoid the impact of inbound censorship, we carefully choose the locations of control servers to deploy them in the countries where no censorship has been widely identified. Besides this, we further conduct experiments to verify that there is no inbound censorship on the control servers' side before conducting measurements. We set up 50 VPN servers from hide-my-ip [24] located in different countries and send a series of probing packets to all of our control servers. Each of these packets carries a test domain in the censored-domain list. We identify that, all of the probing packets with non-censored domains in the corresponding country of the VPN node successfully retrieve the pre-defined static payload from our control servers. This confirms that our control servers have been deployed in the countries where no inbound censorship is enabled.

## 3.4 Ethical Considerations

Ethical concerns in censorship-related measurement studies have been extensively discussed in previous works [30, 31, 44, 55]. In this study, we leverage a RESIP, Proxyrack, as vantage points to issue probing requests, similar to previous studies [30]. Proxyrack is a benefit-driven platform that recruits residential proxies worldwide, where participants can willingly opt-in to join the network and perform tasks for financial profit [1]. We confirmed with Proxyrack that our experiments are within the scope of their Terms of Service, and the participants of Proxyrack RESIP are informed about this type of usage. To reduce the potential risk for the participants of the residential proxies, Pathfinder's design ensures that our experiments do not generate traffic to the actual servers associated with the testing domains so as to avoid access of the actual undesired content. Also, we avoid using the same residential proxy frequently, to minimize the risk of the proxy owners, *i.e.*, each residential IP proxy is only selected once per week to perform our experiments.

Furthermore, we provide a comprehensive description of our experiment in the static payload of our control servers, along with our contact information. Throughout the entire duration of our experiment, we have not received any concerns regarding our measurements and data collection. Finally, measuring censorship activities does not involve human participation or the collection of personal information on any individual or entity, so it is typically outside the scope of the institutional Internal Review Board (IRB) [31]. Despite that, we follow the standard practice of censorship research to obtain an official IRB exemption from our institute.

# 4 EXPERIMENTS

To investigate the censorship inconsistency in real-world scenarios, we conduct two experiments that focus on the primary causes of path diversity that lead to censorship inconsistency: *locations of destinations* and *hosting platforms*. In this section, we provide a detailed introduction to the experiments, including our selection of vantage points and control servers, as well as our data collection processes and results. We also discuss potential ethical issues that may arise from our experiments and our approach to mitigate them.

## 4.1 Impact of IP Destinations

In this experiment, we explore the diverse network paths in packet routing due to different geographically distributed destination addresses. Differing from existing research focusing on network paths induced by router load balancing [9], this experiment is carefully designed so that domain requests originating from a vantage point take various network paths to reach their destinations.

**Vantage Points.** As mentioned in Section 3.2, a vantage point serves as a generic access node in a specific region to measure censorship activities. We use proxies provided by Proxyrack as our vantage points. These proxies are primarily residential, hence traffic from our vantage points simulates real network connections as if it is from regular users.

Moreover, as residential proxies are randomly assigned, one issue we need to address is the uneven distribution of provided proxies across different countries. For instance, we receive a significantly larger number of proxies from several countries including South Korea, Japan, and India, while some countries such as Brazil and South Africa are underrepresented. To rectify this, we establish a cap in Pathfinder, limiting the number of vantage points to a maximum of 80 per country per week. Additional vantage points assigned to us through Proxyrack are discarded, ensuring a more balanced distribution of vantage points for our experiment. This increases the likelihood of observing censorship activities for a larger number of countries while maintaining a substantial amount of data collected from each country.

**Control Servers.** This experiment aims to investigate censorship activities across various network paths. To achieve this, we have carefully selected our control servers based on the following two criteria. First, our control servers must be geographically spread across the world. This increases the likelihood of requests taking diverse network paths to reach these servers. Second, the control servers must be located in places with no known censorship activities, which eliminates the possibility of inbound censorship interfering with our experiment (Section 3.3). Based on these criteria, we establish six control servers using AWS EC2 instances. These servers are located in Virginia (North America - East), California (North America - West), São Paulo (South America), London (Europe), Bahrain (Middle East), and Cape Town (Africa).

**Data Collection.** We conduct our measurement experiments for 60 weeks. Figure 3 shows the distribution of vantage points across all countries. In total, we obtain 144,418 vantage points in 120 countries that have been observed with censorship behaviors in previous studies. Also, these countries represent 6.9 billion (83%) global populations according to the UN population division [16], demonstrating a comprehensive coverage of the majority of Internet users in the world. In addition, the blue line in Figure 5 plots the CDF distribution of the number of vantage points collected in different countries. More than 50% of the countries have vantage points of over 1,500, indicating that we collect a substantial amount of data for each country.

## 4.2 Impact of Hosting Platforms

In addition to the locations of IP destinations, we observe that the hosting platform is another important factor that leads to censorship inconsistency. This is because different peering policies cause packets to be routed through diverse networks that enforce different censorship policies. Thus, we schedule extensive measurements by establishing control servers across different hosting platforms. As such, probing requests from vantage points are routed differently to reach these platforms, which enables us to further evaluate censorship inconsistency caused by various hosting platforms.

**Vantage Points.** We employ the same method described in Section 4.1 to obtain vantage points from a RESIP, Proxyrack. However, each vantage point assigned by Proxyrack has a limited lifespan, which is not sufficient to gather censorship activities for both experiments. Due to this limitation, we use a different set of vantage points for this experiment.

**Control Servers.** In order to gain a comprehensive understanding of the impact of hosting platforms on censorship inconsistency, we set up control servers across three large popular cloud platforms: Amazon Web Services (AWS), Google Cloud Platform (GCP), and Microsoft Azure (Azure). To mitigate the impact of geolocation on censorship inconsistency, we specifically select the locations where all three platforms have data centers deployed. As such, the locations of data centers that we choose to host control servers are Virginia (United States), Sydney (Australia), Paris (France), and São Paulo (South America). With that, we establish one control server in each of these locations from each of the platforms, which forms a set of 12 control servers in total.

**Data Collection.** Our data collection lasts for 36 weeks, during which we measure censorship activities from 66,123 vantage points across 120 countries. Figure 4 presents a heatmap showing the worldwide distribution of these vantage points. Also, the red line in

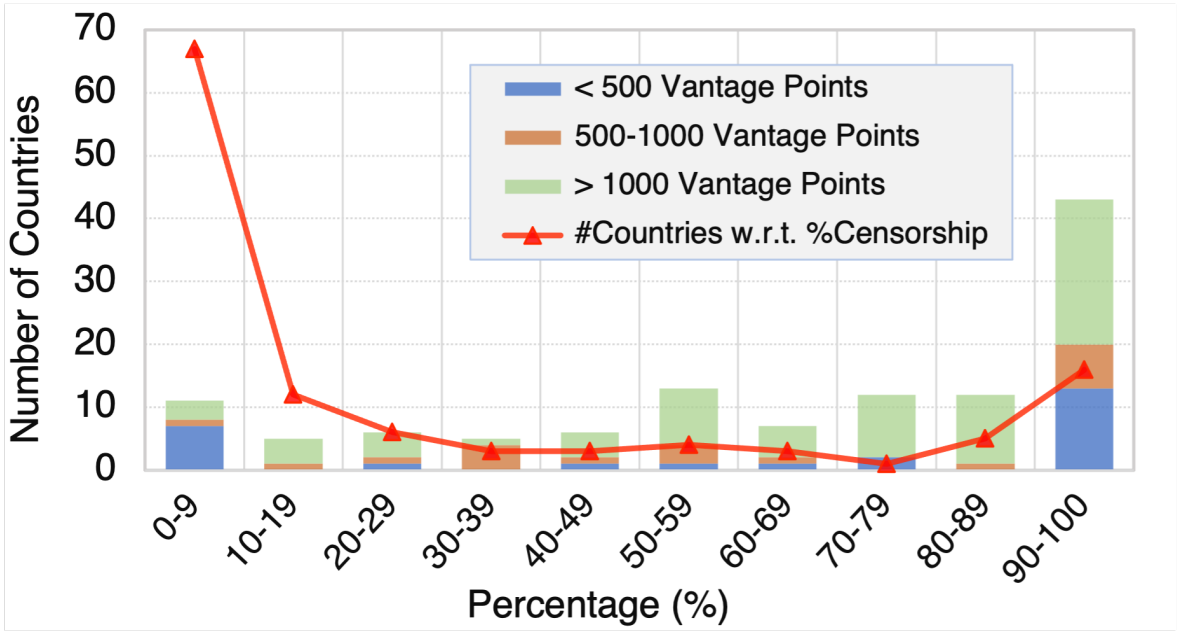

**Figure 6: Distribution of the number of countries with respect to their censorship percentage (*line graph*) and inconsistency percentage (*stacked bar graph*).**

Figure 5 illustrates the CDF distribution of the number of vantage points collected in this experiment, showing that more than 50% of the countries have over 600 vantage points. Although the number of vantage points per country is not as high as that in our IP destination experiment in Section 4.1, we believe that it is still sufficient for a large-scale measurement study.

# 5 RESULTS & ANALYSIS

We here analyze the data collected from experiments in § 4 and answer a series of research questions regarding the existence, cause, and exploitation of censorship inconsistency:

- In general, how prevalent is the censorship inconsistency across different countries and domains? (§ 5.1)
- How extensively do destination servers' locations impact censorship inconsistency in different countries? (§ 5.2)
- How extensively do different hosting platforms lead to censorship inconsistency? (§ 5.3)
- Can specific paths that experience less censorship be utilized to achieve detour for circumvention? (§ 5.4)

## 5.1 Prevalence of Censorship Inconsistency

To answer the first research question, we investigate the prevalence of censorship inconsistency in different countries by analyzing our results from two perspectives: vantage-point-level inconsistency and domain-level inconsistency.

**Vantage-Point-Level Inconsistency.** Pathfinder sends requests from vantage points provided by Proxyrack to our control servers through various network paths, enabling us to identify inconsistent censorship behaviors across different network paths. In this study, we consider a vantage point being censored if any measurement request originating from the vantage point is censored. The censorship percentage is defined as the number of censored vantage points divided by the total number of vantage points being evaluated. More importantly, we define the inconsistency percentage as the percentage of the censored vantage points that observe inconsistent censorship behaviors across different network paths. The censorship percentage implies the extent of censorship deployment and the inconsistency percentage indicates the prevalence of inconsistent censorship behaviors.

| Country | Vantage Points | | | % | % |
|---|---|---|---|---|---|
| | Total | Censored | Incons. | Censored | Incons. |
| Kazakhstan | 1,825 | 1,825 | 894 | 100.00% | 48.99% |
| Kuwait | 1,338 | 1,336 | 1254 | 99.85% | 93.86% |
| China | 1,256 | 1,229 | 1,220 | 97.85% | 99.27% |
| Pakistan | 1,719 | 1,582 | 1,574 | 92.03% | 99.49% |
| Russia | 2,003 | 1,721 | 905 | 85.92% | 52.59% |
| Bangladesh | 1,786 | 1,432 | 1,156 | 80.18% | 80.73% |
| Thailand | 1,925 | 1,182 | 680 | 61.40% | 57.53% |
| Vietnam | 1,971 | 1,106 | 830 | 56.11% | 75.05% |
| India | 2,015 | 1,051 | 992 | 52.16% | 94.39% |
| South Korea | 2,596 | 1,300 | 424 | 50.08% | 32.62% |

**Table 1: List of top 10 countries with more than 1,000 vantage points by the highest censorship percentage and their inconsistency percentage (*Incons.*).**

Figure 6 shows the distribution of the number of countries according to their censorship percentage (the red line). We observe that the majority of countries exhibit an all-or-nothing censorship behavior, similar to the results from Bhaskar *et al.* [9]. Specifically, 79 (66%) countries experience less than 20% censorship percentage, while 21 (17.6%) show more than 80%. Only 20 (16.8%) countries have a censorship percentage between 20% to 80%. These results align with our expectations, as censorship devices are designed to be uniformly deployed across all vantage points.

Figure 6 also presents a bar graph that shows the distribution of the number of countries over censorship inconsistency percentage. Our experiment reveals that only 11 (10%) countries have an inconsistency percentage below 9%, indicating that censorship inconsistency is prevalent worldwide. Moreover, we observe severe censorship inconsistency in a significant number of countries. Specifically, 20 (18%) countries have their inconsistency percentage fall into the range of 90%-99%, and 23 countries even exhibit 100% censorship inconsistency, where all vantage points located in these countries observe various extents of inconsistent censorship among different network paths.

Additionally, the stacked bars in Figure 6 show the different numbers of vantage points we evaluated in our experiment. We collect a substantial amount of data from countries with more than 1,000 vantage points (green bars), allowing us to comprehensively explore censorship behaviors in these countries, by which we observe 34 out of 74 (46%) countries experiencing more than 80% censorship inconsistency. On the other hand, 12 countries with less than 500 vantage points (blue bars) have also shown 100% censorship inconsistency. These results show that our experiments have sufficiently demonstrated the prevalence of censorship inconsistency, even for those counties with fewer vantage points.

We perform further analysis on countries with more than 1,000 vantage points. Table 1 lists the top 10 countries with the highest censorship percentage. We can see that all 10 countries demonstrate censorship percentages exceeding 50%, with Kazakhstan, Kuwait, China, and Pakistan having over 90% vantage points experiencing censorship. Meanwhile, the highlighted countries also exhibit significant censorship inconsistency, for instance, China, Pakistan, and Kuwait show both censorship and inconsistency percentages exceeding 92%. This suggests that, despite implementing strict censorship policies, censorship inconsistency is still prevalent in those countries, resulting in the possibility of evading the censorship by routing through a different path.

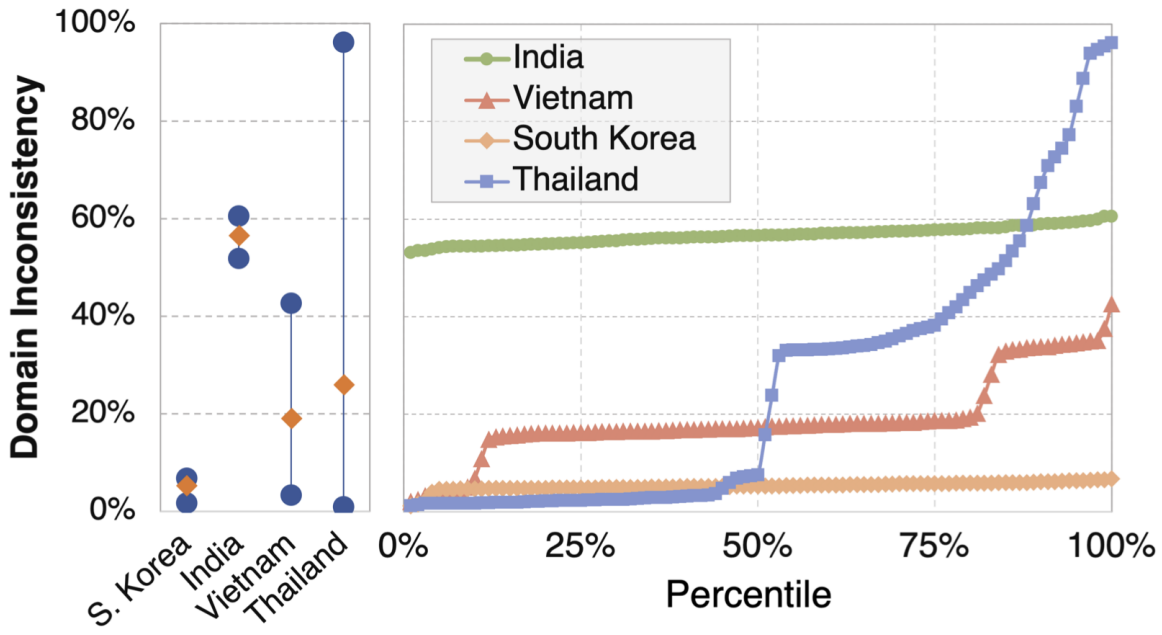

**Figure 7: The min-average-max chart (*left bar graph*) and the percentiles of censorship inconsistency (*right line graph*) in South Korea, India, Vietnam, and Thailand.**

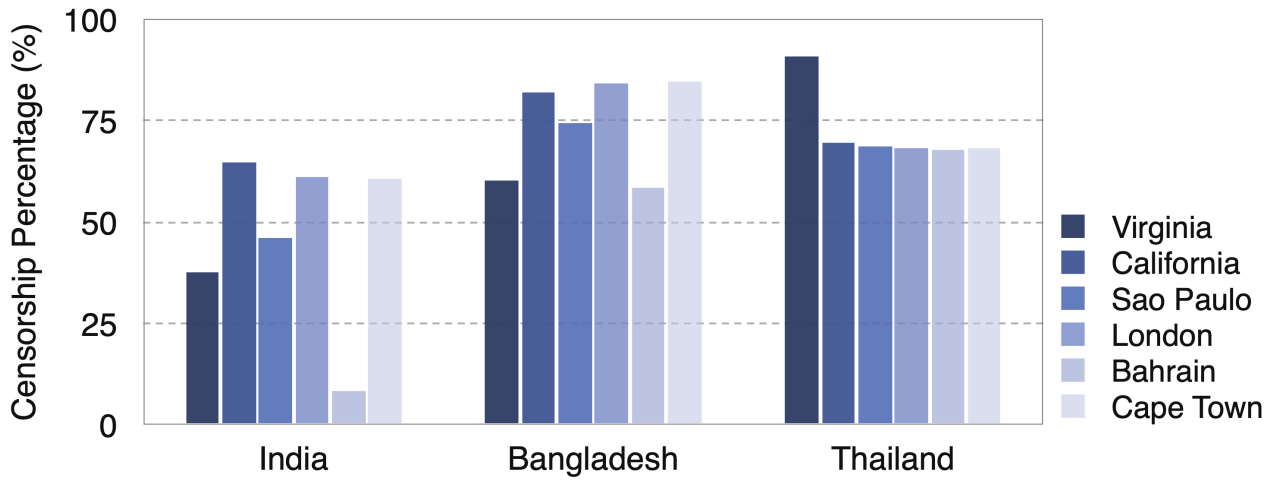

**Figure 8: Censorship percentages to different control servers by VPs in India, Bangladesh, and Thailand.**

**Domain-Level Inconsistency.** In addition to the prevalence of censorship inconsistency encountered by vantage points, we here analyze the inconsistent censorship activities from the perspective of different blocked domains. We consider a domain as being censored if we observe any measurement requests containing that domain name being blocked by censorship devices in a country. We then define a domain-level censorship percentageas the number of probing requests that are blocked by censorship devices divided by the total number of requests we sent in our experiments for a country.

The results show that domain-level censorship inconsistency varies, and is prevalent across many countries. Figure 7 plots the min-average-max chart and the percentiles of domain-level inconsistency in South Korea, India, Vietnam, and Thailand. Specifically, the height of the bar/curve indicates the inconsistency of censorship activities in each country in terms of different domains. We can see that Thailand and Vietnam demonstrate higher-level of domain-level censorship inconsistency, indicating that the censored domains in both countries experience significantly different censorship activities. Among these countries in Figure 7, Thailand shows the most prevalent censorship inconsistency at the domain level. For instance, we observe that a domain, `www.livejasmin.com`, encounters only 1% of censorship activities, whereas 96% of the packets containing `www.bongacams.com` are blocked by censorship, suggesting that both centralized and decentralized censorship may be implemented. Vietnam also demonstrates high censorship inconsistency, with a minimum domain-level censorship of 3% and a maximum of 43%. On the other hand, India and South Korea show a lower censorship inconsistency across different blocked domains, suggesting that each blocked domain in India and South Korea has a similar chance of encountering censorship.

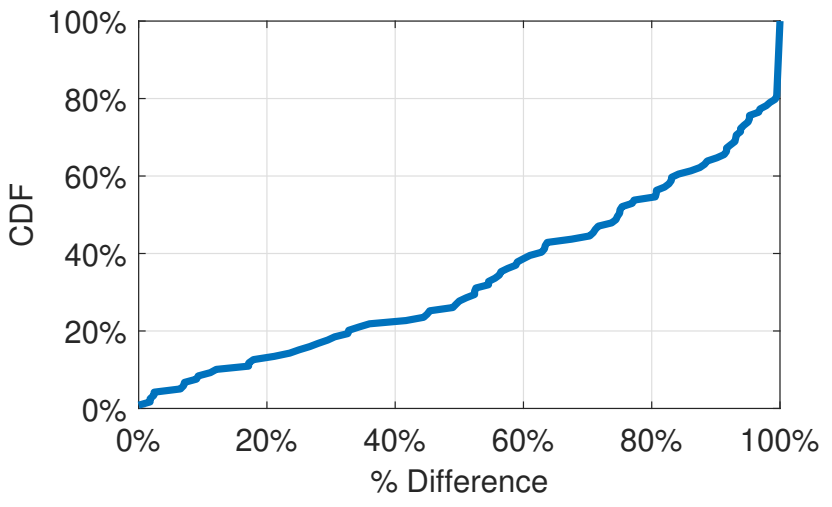

**Figure 9: Distribution of the percentage difference for censorship inconsistency across different countries.**

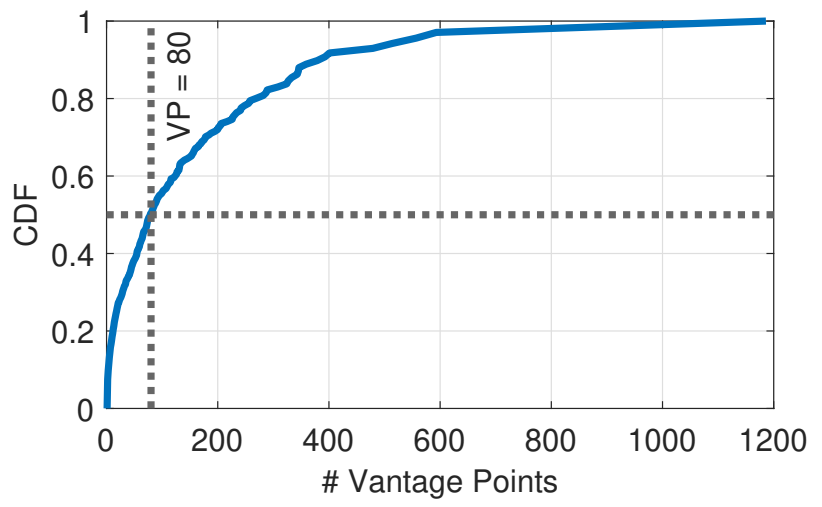

**Figure 10: Distribution of the number of collected vantage points in different ASes.**

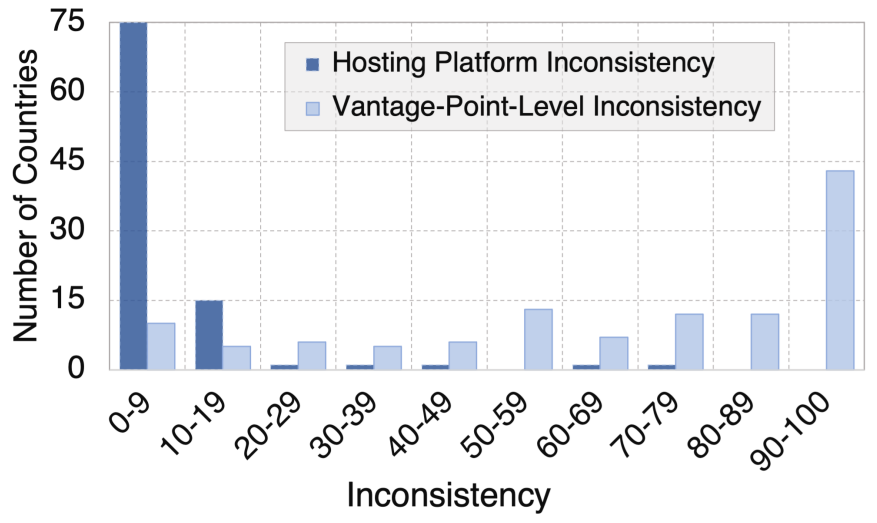

**Figure 11: Distribution of censorship inconsistency in terms of different hosting platforms.**

## 5.2 IP Destination Inconsistency

In this section, we further break down the results presented in Section 5.1 to answer the second research question. As introduced in Section 4.1, we utilize RESIP as vantage points and set up six geographically distributed control servers in various regions of Amazon AWS, enabling our probing requests to traverse different network paths. Then, Pathfinder identifies censorship activities in each network path to examine the impact of the locations of control servers' destinations on censorship inconsistency.

Accordingly, we define the censorship inconsistency caused by IP destinations for each country (*destination inconsistency* for short) as the percentage difference in censorship across different network paths, to further examine the variations in censorship activities across these various network paths. Figure 9 demonstrates the CDF distribution of destination inconsistency in each country. The results show that over 50% of countries experience destination inconsistency greater than 75%, while 20% of countries have an inconsistency of over 30%. This confirms that censorship activities vary widely across network paths toward different control servers.

Figure 8 further illustrates the observed censorship percentages for the network paths toward each control server in three countries, India, Bangladesh, and Thailand. Specifically, we see that India exhibits the most inconsistent censorship activities among various paths. Only 8.4% of the requests toward the control server Bahrain (Middle East) experience censorship, while censorship occurs between 37% to 60% for requests to other control servers. Similarly, Bangladesh's vantage points show the probing requests toward Virginia (North America East) and Bahrain with comparatively less censorship. On the other hand, we observe inconsistent censorship behavior from Thailand where requests sent to Virginia's control server experience significantly higher censorship activities compared to the requests sent to other control servers. These results demonstrate that the location of the destination servers significantly influences the network paths, leading to inconsistent censorship activities.

**AS-level Analysis.** Next, we extend our analysis from the country level to the Autonomous System (AS) level to gain more fine-grained insights into censorship inconsistency. During our experiments, the vantage points are acquired from a total of 3,913 ASes through the RESIP, with a diverse number of obtained vantage points ranging from 1 in AS 3238 and AS 63023 to 1,186 in AS 4766. Figure 10 depicts the distribution of the number of vantage points in different ASes. We filter out the ASes with less than 80 vantage points (represented by the dotted line) in our analysis to focus on those well-represented ASes with sufficient vantage points.

Here, we quantify the AS-level destination inconsistency by measuring the difference between the maximum and minimum censorship percentage encountered by probing packets from an AS sent to different control servers. Table 2 presents the top 10 ASes with the highest censorship inconsistency between different network paths. All ASes in the list experience at least 59.4% inconsistency, with AS 204457 showing the highest censorship inconsistency of 81%. Additionally, these ASes can be categorized into two groups. While AS 204457, AS 137526, and AS 135987 show few paths with a significantly higher censorship percentage than other paths, other ASes have one or two paths with considerably low censorship activities. Importantly, we identify four network paths with no censorship detected, highlighted in Table 2. The observations here at the AS level further strengthen the conclusion that the destination of the packets can lead to noticeable inconsistent censorship, as shown in the country-level analysis in Figure 8.

## 5.3 Hosting Platform Inconsistency

Besides the location of destinations which is one of the primary factors contributing to censorship inconsistency, our experiments also reveal that different hosting platforms (*e.g.*, large cloud providers) also lead to censorship inconsistency since the network paths, through which the request traverses, can be significantly influenced due to the different peering relationships between the hosting platforms and ISPs.

In general, the network paths taken by probing requests are determined by the routing policies and peering relationships established between the ASes from the source to the destination. By setting up control servers in different hosting platforms, the requests would be routed to traverse through diverse network paths due to different connections between upstream ISPs and hosting platforms, enabling us to evaluate the censorship inconsistency resulting from different hosting platforms. To investigate such cases, we set up control servers in the same location but at different cloud platforms (including Amazon AWS, Google Cloud, and Microsoft Azure) as the destinations of probing packets (§ 4.2).

| ASN (Country) | | Censorship Percentage | | | | | | Inconsist. |
|---|---|---|---|---|---|---|---|---|
| | | Virginia | California | São Paulo | London | Bahrain | Cape Town | |
| AS 204457 | (Turkey) | 0.5% | 81.5% | 0.1% | 28.0% | 0.9% | 0.9% | 81.0% |
| AS 137526 | (Bangladesh) | 39.3% | 0.0% | 0.0% | 78.1% | 39.3% | 39.3% | 78.1% |
| AS 135987 | (Vietnam) | 1.2% | 2.4% | 76.8% | 1.2% | 2.4% | 0.0% | 76.8% |
| AS 1312934 | (Thailand) | 69.4% | 9.4% | 1.7% | 71.1% | 71.1% | 70.6% | 69.4% |
| AS 132298 | (Bangladesh) | 67.5% | 67.2% | 66.7% | 68.2% | 0.0% | 65.3% | 68.2% |
| AS 57011 | (Russia) | 76.1% | 0.0% | 67.1% | 67.1% | 67.1% | 67.1% | 67.1% |
| AS 124946 | (India) | 52.6% | 52.8% | 57.8% | 64.9% | 0.0% | 51.9% | 64.9% |
| AS 55492 | (Bangladesh) | 68.6% | 15.5% | 25.9% | 77.3% | 77.7% | 77.7% | 62.1% |
| AS 133227 | (India) | 52.4% | 54.4% | 54.5% | 61.2% | 0.0% | 52.7% | 61.2% |
| AS 199634 | (Russia) | 0.9% | 59.8% | 55.6% | 36.8% | 0.9% | 0.4% | 59.4% |

**Table 2: List of top 10 ASes with the highest censorship inconsistency across various network paths. (***Inconsist.: Inconsistency, measured by the greatest difference of AS-level censorship percentages***)**

| Country | # of Requests | % Censorship | | | Inconsist. |
|---|---|---|---|---|---|
| | | AWS | GCP | Azure | |
| South Korea | 763,875 | 98.93% | **32.21%** | 76.64% | 66.72% |
| Thailand | 90,962 | 92.14% | 85.39% | 44.11% | 48.04% |
| India | 405,281 | 50.89% | **18.57%** | 37.95% | 32.32% |
| Russia | 1,265,372 | 88.41% | 77.62% | 89.01% | 11.39% |
| Venezuela | 104,367 | 19.71% | 22.15% | 30.83% | 11.12% |
| Jamaica | 84,275 | 2.03% | 1.62% | 12.19% | 10.57% |
| Poland | 103,476 | 0.82% | 0.82% | 10.75% | 9.93% |
| South Africa | 91,285 | 3.19% | 2.91% | 12.76% | 9.85% |
| Canada | 128,472 | 7.04% | 7.54% | 16.48% | 9.43% |
| Brazil | 130,232 | 1.02% | 0.91% | 10.00% | 9.09% |

**Table 3: Top 10 countries with more than 800 vantage points and their hosting platform inconsistency.**

Similar to Section 5.2, we then define censorship inconsistency caused by hosting platforms (*hosting inconsistency* for short) as the percentage difference in censorship encountered by probing packets sent to various hosting platforms. Figure 11 illustrates the distribution of the number of countries with respect to such hosting platform inconsistency (shown as dark blue). The results reveal that 16 (13.3%) countries have censorship inconsistency of more than 10%, indicating that inconsistent censorship behaviors toward various hosting platforms largely exist in small number of countries. Furthermore, compared to the vantage-point-level inconsistency from Section 5.1 (*i.e.*, the bar graph in Figure 6, shown as light blue in Figure 11), the hosting platform inconsistency does not show any apparent correlation. This indicates that our observation here is indeed the impact of hosting platforms, not being coincidentally affected by the destinations of control servers in different clouds.

Furthermore, we conduct additional analysis focusing on the most well-represented countries studied through our experiments. Table 3 shows the top 10 countries with at least 800 acquired vantage points. We can see that the censorship inconsistency with respect to different hosting platforms ranges from 9.09% to 66.72%. In particular, South Korea demonstrates significantly lower censorship in GCP with only 32.21% (marked in red in Table 3). In contrast, network paths toward AWS and Azure demonstrate 98.93% and 76.64% censorship percentages, respectively. India shows a similar trend, with paths to GCP showing significantly lower censorship (18.57%) than the other hosting platforms (50.89% in AWS and 37.95% in Azure). Both countries are highlighted in Table 3, and we further conduct detailed case studies for these two countries through application traceroute to investigate the root cause of this censorship inconsistency, which we elaborate on in Section 6.

Table 3 illustrates inconsistent censorship caused by different cloud platforms. To further break down the results into different control servers hosted in each cloud, we examine the censorship percentages associated with each individual control server. Figure 12 illustrates the results of three countries, South Korea, Thailand, and India, comparing the censorship percentage observed on the paths toward each control server hosted on GCP, AWS, and Azure. The figure shows that the censorship percentage varies significantly on each hosting platform, confirming the aggregated results at the hosting platform level. Moreover, we can see that censorship percentages for all three countries are relatively stable with GCP and Azure, indicating that (1) the location of our control servers does not have a significant impact on the censorship percentage in these two platforms and (2) GCP and Azure may establish more consistent connections with the networks in these countries. On the other hand, network paths to AWS, specifically for South Korea and India, show a more inconsistent censorship percentage among different locations, implying that its peering connections may be more diverse. This also aligns with our discussion about the impact of IP destinations on censorship inconsistency in § 5.2.

## 5.4 Potential of Censorship Circumvention

Censorship inconsistency provides users with a potential strategy to evade censorship and gain access to censored content. Previous sections have discussed that the location of the destination and the hosting platforms can significantly impact the routing paths of requests, implying that censorship circumvention can be potentially achieved by the careful selection of outbound network paths.

To evade censorship for certain domains, one straightforward strategy can be to deliberately route the requests to proxy servers located where less/no censorship occurs in network paths. For example, as shown in Figure 8, vantage points in India experience fewer censorship activities towards the control server located in Bahrain (Middle East). Hence, selecting proxy servers in Bahrain may unblock the majority of the requests coming from India. On the other hand, Thailand demonstrates that requests toward the control server located in Virginia would experience more aggressive censorship, hence avoiding routing requests to this location would be a desired option in terms of circumvention.

In addition to the location of destinations, censorship could also be evaded with the careful selection of the hosting platforms of proxy servers. As highlighted in Table 3, vantage points from both South Korea and India observe significantly low censorship activities in the network paths toward GCP. With that, it is preferred to deliberately select proxy servers hosted on GCP to evade censorship. Also, service providers could intentionally host their services on less-censored platforms, ensuring that users have a better chance to access their services with less interference.

Moreover, such strategies could be integrated with existing circumvention frameworks, to actively and dynamically explore potential circumventing paths as an additional vector in circumvention toolkit. They could also be combined with other methods, *e.g.*, CenFuzz [50], Geneva [11], and NetShuffle [35], to create a more efficient and robust, yet still simple method to achieve censorship circumvention.

# 6 APPLICATION TRACEROUTE: CASE STUDIES

As shown in § 5.3, the hosting platform could largely affect the severity of censorship activities experienced by users from certain countries. For instance, we observed that South Korea and India show lower censorship activities in network paths toward Google Cloud Platform (GCP) than Amazon AWS and Microsoft Azure. However, the RESIP vantage points leveraged by Pathfinder cannot perform traceroute to pinpoint the exact location on a network path where censorship occurs, because it is not feasible to set the TTL values of the relayed packets due to the fact that RESIPs usually operate above the transport layer. Therefore, in this section, we leverage application traceroute with commercial VPNs as vantage points to conduct a detailed, end-to-end investigation for censorship inconsistency in the above two countries.

## 6.1 Methodology

As described in § 2.2, application traceroute works by sending a series of requests with incremental TTLs and is accomplished when receiving a sign of censorship or our pre-defined static payload. By examining the corresponding response of each request, we are able to reconstruct the network path taken by the probing requests and identify the specific hop where the censorship device is deployed. To collect such path information, we rely on commercial VPNs to perform the application traceroute experiments.

A prior study [59] reveals that many VPN services may lie about their server locations. As the authenticity of the location of vantage points are important to our analysis, we validate whether the VPN servers are located where they advertise. To achieve this, we first attempt to trigger censorship from a VPN server by accessing some censored domains. We then confirm the location of the VPN servers if we receive official blockpages established by the censoring countries (*i.e.*, South Korean or Indian government in our case studies, see Section 6.2 and 6.3). By examining several different VPN services, we choose hide-my-ip [24]'s VPN as its servers advertises relatively reliable location information.

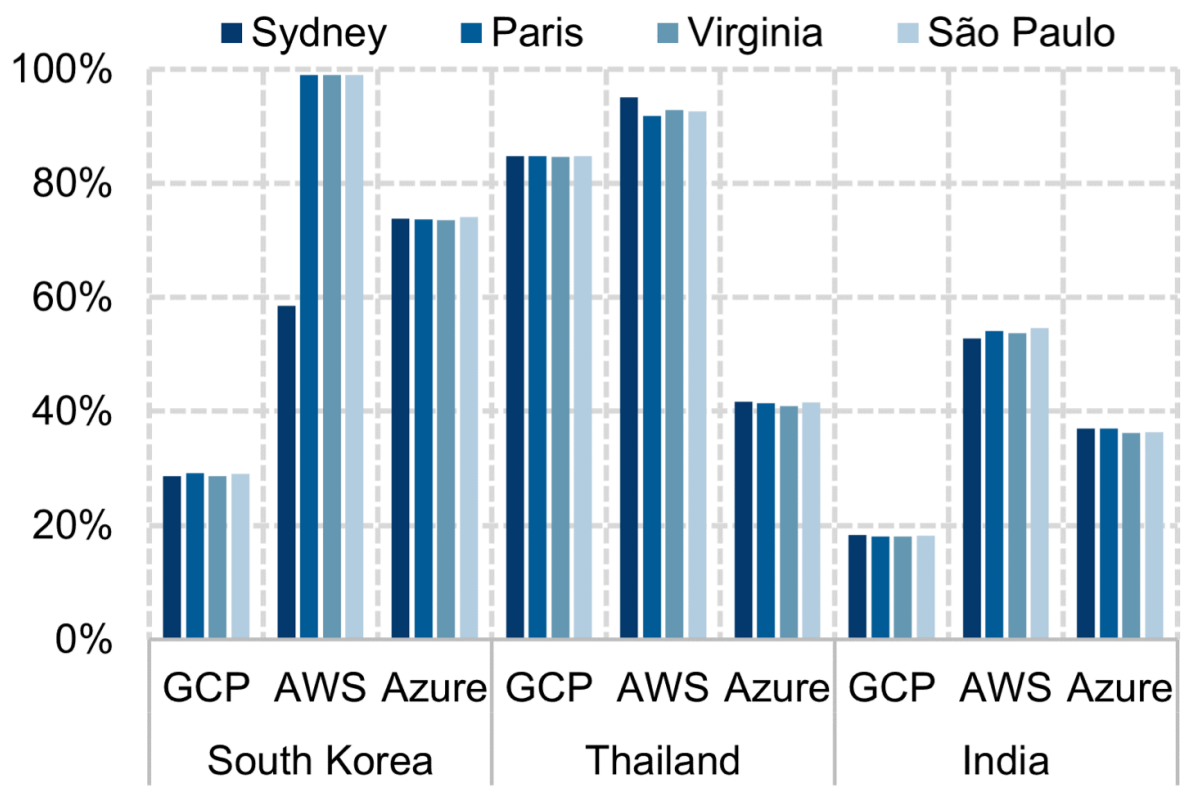

**Figure 12: Censorship percentages from South Korea, Japan, & India to control servers in GCP, AWS, & Azure.**

We use the same set of control servers introduced in § 4.2 as destinations for the application traceroute. These servers are hosted by various cloud providers (Amazon AWS, Google Cloud, and Microsoft Azure) and are geographically located across the globe (Sydney, Paris, Virginia, and São Paulo). For both studied countries (South Korea and India), we randomly select 32 domains from the censored domains list (§ 3.2) and initiate the application traceroute carrying these domains from each of the VPN servers to each of the control servers.

## 6.2 Case Study: South Korea

Leveraging the application traceroute, we are able to identify the network devices on the paths. Figure 13 shows a concrete example of the paths taken by the probing requests between our vantage points in Seoul, South Korea, and distributed control servers around the world. We observe that censorship consistently occurs on the paths to three control servers hosted by Amazon AWS. Specifically, we receive an official blockpage (through the redirection to `http://warning.or.kr`) indicating that the accessed domain is prohibited. The remaining paths, including one path to Amazon AWS in Sydney, did not encounter censorship activities. This aligns with our general observations in Figure 12, where AWS shows more diverse censorship behaviors and the paths to AWS's Sydney site experience less censorship. This could be caused by the censorship devices on this path applying a different domain list that is less aggressive than others.

On the other hand, we observe no censorship occurring on the paths toward Google Cloud and Microsoft Azure. Although we observe that the probing packets, including those to AWS, are all routed with the same ISP (AS 4766, Korea Telecom), the paths are different from certain hops, which leads to inconsistent censorship. More interestingly, we observe the upstream ISP (Korea Telecom) peered with GCP's private networks within South Korea, resulting in the probing requests entering the private networks of GCP before reaching the censorship devices. Table 5 in Appendix B provides more detailed application traceroute results for one domain, including all routers' ASes and ISPs on the paths.

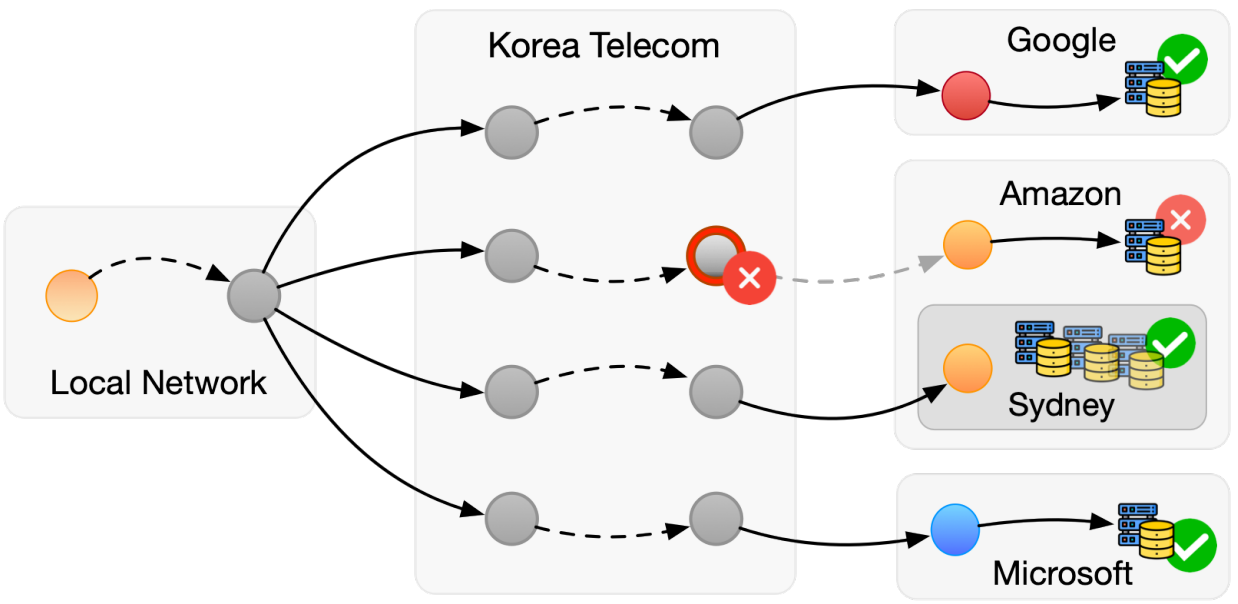

**Figure 13: Application traceroute of different paths over different hosting platforms in Seoul, South Korea.**

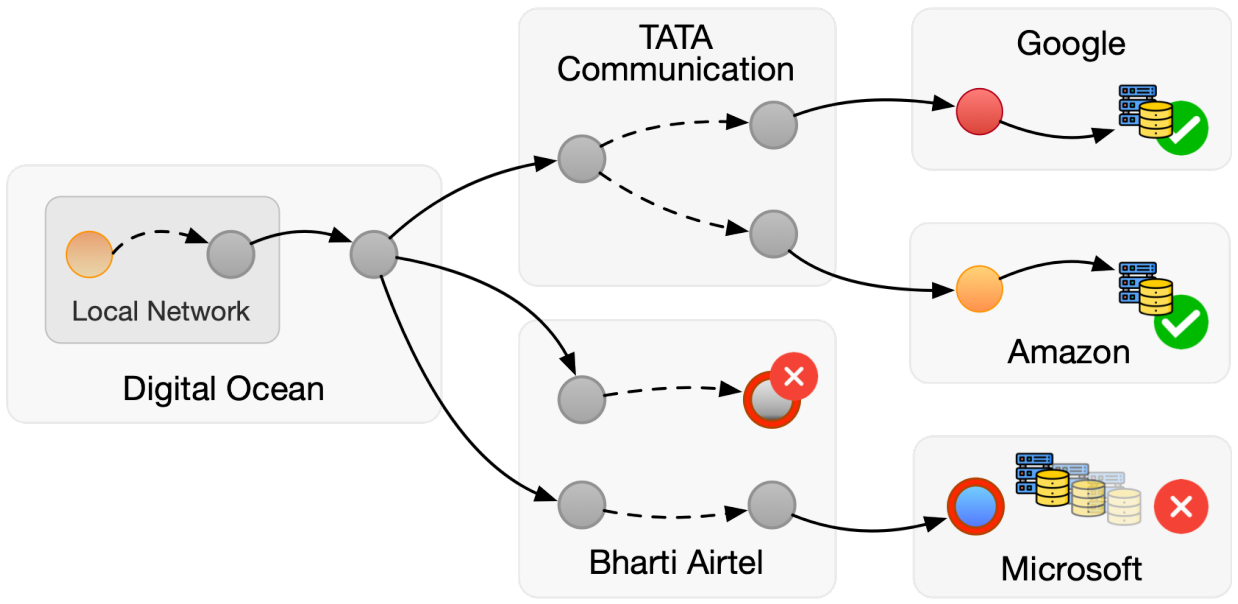

**Figure 14: Application traceroute of different paths over different hosting platforms in Bangalore, India.**

## 6.3 Case Study: India

Next, we conduct a case study with application traceroute in India and explore censorship inconsistency in depth. Figure 14 depicts one case where the hosting platforms play a key role in inconsistent censorship behaviors. Our experiments reveal that censorship always occurs on the paths to Microsoft Azure by receiving an official block page. However, we did not encounter any censorship when retrieving static payloads from control servers on AWS and GCP.

Table 6 in Appendix B displays a detailed illustration of application traceroute performed in Bangalore, India. Through the traceroute results, we observe specific censorship devices deployed in the ASes including AS 9498 (Bharti Airtel) and AS 8057 (Microsoft Azure). However, as censorship is not likely to be enabled by Microsoft's public cloud, we consider that the censorship devices should still be located at AS 9498 (Bharti Airtel). We speculate this experiment results could be interfered with by some censorship devices copying TTL values from the original probing packet, resulting in increased TTL values when tracing the censor's location. Such TTL-copying behavior by censorship devices has been detected and extensively examined in prior study [30].

On the other hand, the packets sent to GCP and AWS are both routed through AS 4755 (TATA Communication) and experience no censorship. Furthermore, similar to the case in South Korea, we observe that no censorship occurred on the paths to the control servers on GCP while no intermediate hops are visible before reaching the control servers, implying that the probing packets are routed to GCP's private network by a direct peer between GCP and AS 4755.

Comparatively, the results of the application traceroute in South Korea show all probing requests to different clouds passed through one upstream provider (AS 4766) before encountering any censorship. In India, the requests issued from one vantage point are routed through two different providers, where one of the provider's networks (AS 9498) enables censorship for the test domain. As shown in Table 3, both South Korea and India experience less censorship on paths toward GCP's control servers. This is due to the peering between local ISPs and GCP's private networks, resulting in no censorship on paths toward GCP's control servers.

## 7 LIMITATIONS & FUTURE WORK

**Censorship Close to Vantage Points.** Pathfinder aims to examine censorship deployments by varying probing packets' destinations to explore diverse paths. As shown in our experimental results, it can largely identify different censorship implementations in the various upstream transit networks. However, although censorship devices are more commonly deployed close to a nation's border networks [30], censorship enforced at the hops close to clients also exists [50] but may not be detected by Pathfinder. In such a case, the vantage points would observe consistent censorship behaviors because all probing packets are routed by the same ISP.

**Vantage Point Selection.** In this study, we utilize the RESIP provided by Proxyrack [45] as vantage points to investigate inconsistent censorship activities in various network paths. We obtain a total of 144K and 66K vantage points from 120 countries in our IP destination and hosting platform experiments, respectively. While many countries are extensively studied with a large number of vantage points, other countries, such as Switzerland, Sudan, and Zambia, are only collected with less than 200 vantage points due to limited available residential nodes. Also, to examine the censorship inconsistency in more detail, we performed case studies using application traceroute through VPNs, instead of RESIPs, because RESIPs typically do not support changing packets' TTLs for application traceroute. However, although VPNs are also widely used in existing studies [40, 48, 50], VPN servers are commonly hosted in commercial data centers, where the traffic may encounter less censorship than in residential networks. These vantage point biases could be rectified by recruiting more vantage points from different platforms, but the skewness of node distribution exists for all these kinds of measurements.

**Test Domain List.** Our experiments utilize the existing censored domain list collected from Disguiser [30] to reduce measurement efforts for identifying censored domains. While the domains on the list have been validated with their censorship status, this previous study does not consider the impact of censorship inconsistency, so it may miss some censored domains if the probing requests encounter no censorship on certain testing network paths. However, this should not impact our study much because an incomplete list of censored domains is still sufficient to explore diverse network paths and illustrate the censorship deployments.

**Future Work.** In this study, we primarily focus on using HTTP-based censorship as a lens to explore and understand inconsistent censorship deployments at scale. In the future, we will further extend the experiments to DNS-based and HTTPS/SNI-based censorship for characterizing and comparing their operations and behaviors. Also, we will continuously run Pathfinder to collect longitudinal data and identify the changes of the extent of censorship inconsistency across different countries.

## 8 RELATED WORK

### 8.1 Global-Scale Censorship Measurements

OONI (Open Observatory of Network Interference) [21] has established a community-driven global measurement framework by recruiting participants to run pre-defined measurements to investigate censorship activities. VabderSloot *et al.* [55] proposed Quack, a remote measurement system that efficiently explores application-layer interference by using the Echo protocol. Built upon Quack, FilterMap [49] improved Quack to identify the content filtering techniques by analyzing their block pages. Niaki *et al.* [40] developed ICLab, a platform that employs VPNs as vantage points to launch a variety of longitudinal censorship measurements and proposed techniques to identify unknown blockpages. Censored Planet [48] integrates multiple existing techniques/frameworks and enables synchronized censorship measurements to enhance data representativeness and coverage. In addition, Pearce *et al.* [44] introduced Iris, a system designed to identify and characterize DNS censorship on a global scale. Then, Bhaskar *et al.* [9] further presented BreadCrumb, a tool that measures DNS censorship variation by manipulating the source parameters of probe packets and router-based load balancing. Lastly, Jin *et al.* [30] presented Disguiser, an end-to-end, ground truth based measurement platform that detects censorship activities and reveals the censor deployment.

Our study complements existing research efforts by designing Pathfinder to systematically investigate censorship activities in various network paths on a global scale, revealing the wide existence of censorship inconsistency.

### 8.2 Country-Specific Censorship Studies

While significant research efforts on censorship studies have been undertaken on a global scale, many previous research studies have also been devoted to investigating censorship behaviors in specific countries. As one of the largest censorship systems in the world, the Great Firewall of China (GFW) has been extensively studied [5–7]. Xu *et al.* [61] examined China's border ASes networks and its relationship with foreign countries. Ensafi *et al.* [17] measured the reachability of servers and networks blocked by the GFW using connectivity measurement techniques. Hoang *et al.* [25] developed GFWatch to examine the DNS filtering behaviors of the GFW. Besides GFW, other research studies have focused on Internet censorship deployed by Iran [8, 10], Pakistan [34, 38], Syria [13], India [62], Kazakhstan [47], and Russia [51].

Although similar inconsistent or decentralized censorship activities have been witnessed in several country-specific studies, their observations typically rely on the collaboration from activists on the ground, and the used methods/data cannot be extended to other countries. Instead, Pathfinder aims to systematically examine the censorship inconsistency on a global scale by exploring diverse network paths.

### 8.3 Censorship Circumvention

With the ever-increasing censorship activities on the Internet, censorship circumvention techniques have also been widely examined [53]. Fifield *et al.* [20] proposed Domain Fronting, which evades censorship detection by concealing the domain names of the communication partners. This technique has also been further explored by many other circumvention systems such as Lantern [36] and Psiphon [46]. Domain Shadowing [57] exploits the fact that CDNs allow their customers to claim arbitrary domains as the back-end, which can be set as blocked domains to evade censorship. Burnett *et al.* [12] developed Collage, enabling users to exchange messages through cover traffic in sites that host user-generated content. Khattak *et al.* [33] introduced an analysis model that inspects the evasion vulnerabilities discovered by Network Intrusion Detection System (NIDS). Autosonda [28] is a tool designed to study web filters and discover a range of implementation and decision-making techniques. Nisar *et al.* [41] proposed C-Saw, a circumvention system that integrates censorship measurements with circumvention techniques into a single system. Bock *et al.* [11] proposed Geneva, a genetic algorithm that automates the discovery of censorship circumvention strategies against on-path network censors. Wang *et al.* [56] exploited the discrepancies in TCP state machines of deep packet inspection (DPI) implementation to bypass censorship. Raman *et al.* [50] developed Cenfuzz, which employs different HTTP methods to circumvent censorship devices and identify the evasion behavior of vendors by clustering. Other circumvention tools include CDNBrowsing systems [26, 39, 63], Decoy routing [27, 32], Flash proxy [19], Infranet [18], Telex [60], uProxy [54], Alkasir [3], LASTor [2], Astoria [42], *etc.*

Our study uncovers censorship inconsistencies, which can also be leveraged as a complementary component to be integrated with existing censorship circumvention (§ 5.4).

## 9 CONCLUSION

Internet censorship is the control or suppression mechanism on what online content users can access. In this study, we designed and implemented Pathfinder, a framework for investigating inconsistent censorship activities in different network paths inside a country. Our findings reveal the prevalent existence of censorship inconsistency in many countries due to the changes of network paths. We further demonstrated that geolocation and hosting platforms of destination servers often result in the requests being routed through various network paths and encountering inconsistent censorship activities. We showed that such censorship inconsistency can be exploited to circumvent censorship in many countries. To further investigate censorship inconsistency, we leveraged the application traceroute to identify the exact node in a network path where such inconsistency occurs.

## A OBSERVED CENSORSHIP INCONSISTENCY IN COUNTRIES

| Country | Percentage | # of Countries |
|---|---|---|
| Albania, Angola, Bolivia, Bosnia Herzegovina, Brazil, Gambia, Honduras Jamaica, Madagascar, Malawi, Mauritius, Myanmar, Nepal, New Caledonia Oman, Nepal, Puerto Rico, Somalia, Tajikistan, Uruguay | 100% | 19 |
| Algeria, Bahrain, China, Estonia, Germany, Greece India, Indonesia, Japan, Kenya, Kuwait, Libya, Mexico Netherlands, Pakistan, Portugal, Serbia, Sweden | 90% - 99% | 18 |
| Argentina, Australia, Bangladesh, Belgium, Czechia, Georgia Italy, Norway, Spain, Suriname, United Kingdom, Venezuela | 80% - 89% | 12 |
| Belarus, Chile, France, Hong Kong, Laos, Lebanon, New Zealand Philippines, Singapore, South Africa, Uganda, Vietnam | 70% - 79% | 12 |
| Canada, Hungary, Israel, Lithuania, Macao, Palestine, Palestinian Territory | 60% - 69% | 7 |
| Benin, Colombia, Ecuador, El Salvador, French Polynesia, Iran, Latvia Nigeria, Panama, Taiwan, Thailand, Ukraine | 50% - 59% | 13 |
| Afghanistan, Egypt, Kazakhstan, Morocco, Nicaragua, Turkey | 40% - 49% | 6 |
| Curacao, Dominican Republic, Finland, Jordan, Malaysia, Moldova Qatar, South Korea, SriLanka, United Arab Emirates, Yemen | 20% - 39% | 11 |
| Azerbaijan, Slovenia, Bulgaria, Uzbekistan, Saudi Arabia, Poland Cuba, Peru, Fiji, Brunei, Equatorial Guinea, Ivory Coast | 0% - 19% | 12 |

**Table 4: Fraction of VPs that Observe Censorship Inconsistency in Censored Countries/Regions > 200**

## B APPLICATION TRACEROUTE RESULTS

Table 5 displays a detailed application traceroute result that depicts the paths taken by probing requests from one vantage point located in Seoul, South Korea, to 12 control servers deployed in three clouds for the censored domain `torrentdada.com`.

Table 6 presents the application traceroute from one vantage point located in Bangalore, India, to the control servers for the censored domain `cckerala.com`.

Control servers are located in Sydney, Paris, Virginia, and São Paulo and are hosted on three cloud platforms: Amazon AWS, Google Cloud Platform, and Microsoft Azure. At each hop in the traceroute, we show the corresponding IPs and associated ASes. We have also highlighted the censors in red to indicate the censorship occurrences to demonstrate the inconsistent activities identified on different paths.

| Hops | Traceroutes to Control Servers | | | | | | | | | | | |
|---|---|---|---|---|---|---|---|---|---|---|---|---|
| | Amazon AWS | | | | Google Cloud | | | | Microsoft Azure | | | |
| | Sydney | Paris | Virginia | São Paulo | Sydney | Paris | Virginia | São Paulo | Sydney | Paris | Virginia | São Paulo |
| ttl = 1 | 10.238.0.1<br>local network | 10.238.0.1<br>local network | 10.238.0.1<br>local network | 10.238.0.1<br>local network | 10.238.0.1<br>local network | 10.238.0.1<br>local network | 10.238.0.1<br>local network | 10.238.0.1<br>local network | 10.238.0.1<br>local network | 10.238.0.1<br>local network | 10.238.0.1<br>local network | 10.238.0.1<br>local network |
| ttl = 2 | 172.32.2.10<br>local network | 172.32.2.10<br>local network | 172.32.2.10<br>local network | 172.32.2.10<br>local network | 172.32.2.10<br>local network | 172.32.2.10<br>local network | 172.32.2.10<br>local network | 172.32.2.10<br>local network | 172.32.2.10<br>local network | 172.32.2.10<br>local network | 172.32.2.10<br>local network | 172.32.2.10<br>local network |
| ttl = 3 | 172.30.10.17<br>local network | 172.30.10.17<br>local network | 172.30.10.17<br>local network | 172.30.10.17<br>local network | 172.30.10.17<br>local network | 172.30.10.17<br>local network | 172.30.10.17<br>local network | 172.30.10.17<br>local network | 172.30.10.17<br>local network | 172.30.10.17<br>local network | 172.30.10.17<br>local network | 172.30.10.17<br>local network |
| ttl = 4 | 172.30.10.9<br>local network | 172.30.10.9<br>local network | 172.30.10.9<br>local network | 172.30.10.9<br>local network | 172.30.10.9<br>local network | 172.30.10.9<br>local network | 172.30.10.9<br>local network | 172.30.10.9<br>local network | 172.30.10.9<br>local network | 172.30.10.9<br>local network | 172.30.10.9<br>local network | 172.30.10.9<br>local network |
| ttl = 5 | 119.196.0.77<br>AS4766<br>Korea Telecom | 119.196.0.77<br>AS4766<br>Korea Telecom | 119.196.0.77<br>AS4766<br>Korea Telecom | 119.196.0.77<br>AS4766<br>Korea Telecom | 119.196.0.77<br>AS4766<br>Korea Telecom | 119.196.0.77<br>AS4766<br>Korea Telecom | 119.196.0.77<br>AS4766<br>Korea Telecom | 119.196.0.77<br>AS4766<br>Korea Telecom | 119.196.0.77<br>AS4766<br>Korea Telecom | 119.196.0.77<br>AS4766<br>Korea Telecom | 119.196.0.77<br>AS4766<br>Korea Telecom | 119.196.0.77<br>AS4766<br>Korea Telecom |
| ttl = 6 | 112.190.32.21<br>AS4766<br>Korea Telecom | 112.190.32.21<br>AS4766<br>Korea Telecom | 112.190.32.21<br>AS4766<br>Korea Telecom | 112.190.32.21<br>AS4766<br>Korea Telecom | 112.190.32.21<br>AS4766<br>Korea Telecom | 112.190.32.21<br>AS4766<br>Korea Telecom | 112.190.32.21<br>AS4766<br>Korea Telecom | 112.190.32.21<br>AS4766<br>Korea Telecom | 112.190.32.21<br>AS4766<br>Korea Telecom | 112.190.32.21<br>AS4766<br>Korea Telecom | 112.190.32.21<br>AS4766<br>Korea Telecom | 112.190.32.21<br>AS4766<br>Korea Telecom |
| ttl = 7 | * | * | * | * | * | * | * | * | * | * | * | * |
| ttl = 8 | * | 112.174.90.110<br>AS4766<br>Korea Telecom | 112.174.90.226<br>AS4766<br>Korea Telecom | 112.174.90.154<br>AS4766<br>Korea Telecom | * | * | * | * | * | * | * | * |
| ttl = 9 | 112.191.117.101<br>AS4766<br>Korea Telecom | **Censor**:<br>112.174.91.218<br>AS4766<br>Korea Telecom | **Censor**:<br>112.174.91.182<br>AS4766<br>Korea Telecom | **Censor**:<br>112.174.86.238<br>AS4766<br>Korea Telecom | 128.134.10.246<br>AS4766<br>Korea Telecom | 128.134.10.246<br>AS4766<br>Korea Telecom | 128.134.10.246<br>AS4766<br>Korea Telecom | 128.134.10.246<br>AS4766<br>Korea Telecom | 121.189.3.138<br>AS4766<br>Korea Telecom | 121.189.3.138<br>AS4766<br>Korea Telecom | 121.189.3.138<br>AS4766<br>Korea Telecom | 121.189.3.138<br>AS4766<br>Korea Telecom |
| ttl = 10 | 112.191.118.177<br>AS4766<br>Korea Telecom | | | | 34.151.125.165<br>(Sydney)<br>Google Cloud | 34.163.60.19<br>(Paris)<br>Google Cloud | 35.245.157.97<br>(Virginia)<br>Google Cloud | 35.247.224.42<br>(São Paulo)<br>Google Cloud | 104.44.239.244<br>AS8075<br>Microsoft Corp. | 104.44.239.240<br>AS8075<br>Microsoft Corp. | 104.44.239.244<br>AS8075<br>Microsoft Corp. | 104.44.239.244<br>AS8075<br>Microsoft Corp. |
| ttl = 11 | 211.47.31.78<br>AS4766<br>Korea Telecom | | | | | | | | 104.44.22.41<br>AS8075<br>Microsoft Corp. | 104.44.22.21<br>AS8075<br>Microsoft Corp. | 104.44.22.41<br>AS8075<br>Microsoft Corp. | 104.44.22.49<br>AS8075<br>Microsoft Corp. |
| ttl = 12 | 150.222.116.153<br>Amazon Tech.<br>KR Seoul | | | | | | | | 104.44.19.241<br>AS8075<br>Microsoft Corp. | 104.44.30.15<br>AS8075<br>Microsoft Corp. | 104.44.19.241<br>AS8075<br>Microsoft Corp. | 104.44.28.247<br>AS8075<br>Microsoft Corp. |
| ttl = 13 | 54.239.123.133<br>AS16509<br>Amazon.com<br>KR Seoul | | | | | | | | 104.44.17.109<br>AS8075<br>Microsoft Corp. | 104.44.30.16<br>AS8075<br>Microsoft Corp. | 104.44.17.109<br>AS8075<br>Microsoft Corp. | 104.44.16.241<br>AS8075<br>Microsoft Corp. |
| ttl = 14 | * | | | | | | | | 104.44.7.231<br>AS8075<br>Microsoft Corp. | 104.44.7.13<br>AS8075<br>Microsoft Corp. | 104.44.7.231<br>AS8075<br>Microsoft Corp. | 104.44.19.112<br>AS8075<br>Microsoft Corp. |
| ttl = 15 | 15.230.212.61<br>Amazon Tech.<br>Australia Sydney | | | | | | | | 104.44.19.112<br>AS8075<br>Microsoft Corp. | 104.44.17.67<br>AS8075<br>Microsoft Corp. | 104.44.17.182<br>AS8075<br>Microsoft Corp. | 104.44.17.182<br>AS8075<br>Microsoft Corp. |
| ttl = 16 | 15.230.210.56<br>Amazon Tech.<br>Australia Sydney | | | | | | | | 104.44.17.203<br>AS8075<br>Microsoft Corp. | 104.44.29.50<br>AS8075<br>Microsoft Corp. | 104.44.17.182<br>AS8075<br>Microsoft Corp. | 104.44.19.85<br>AS8075<br>Microsoft Corp. |
| ttl = 17 | 15.230.210.91<br>Amazon Tech.<br>Australia Sydney | | | | | | | | * | 104.44.11.226<br>AS8075<br>Microsoft Corp. | 104.44.30.191<br>AS8075<br>Microsoft Corp. | 104.44.16.5<br>AS8075<br>Microsoft Corp. |
| ttl = 18 | 15.230.210.152<br>Amazon Tech.<br>Australia Sydney | | | | | | | | 104.44.7.123<br>AS8075<br>Microsoft Corp. | * | 104.44.30.191<br>AS8075<br>Microsoft Corp. | 104.44.30.107<br>AS8075<br>Microsoft Corp. |
| ttl = 19 | 15.230.211.34<br>Amazon Tech.<br>Australia Sydney | | | | | | | | * | * | 104.44.16.161<br>AS8075<br>Microsoft Corp. | 104.44.28.254<br>AS8075<br>Microsoft Corp. |
| ttl = 20 | * | | | | | | | | * | * | 104.44.28.254<br>AS8075<br>Microsoft Corp. | 104.44.28.223<br>AS8075<br>Microsoft Corp. |
| ttl = 21 | * | | | | | | | | * | * | 104.44.28.223<br>AS8075<br>Microsoft Corp. | 104.44.29.44<br>AS8075<br>Microsoft Corp. |
| ttl = 22 | * | | | | | | | | * | * | 104.44.29.44<br>AS8075<br>Microsoft Corp. | 104.44.17.216<br>AS8075<br>Microsoft Corp. |
| ttl = 23 | * | | | | | | | | * | * | 104.44.28.120<br>AS8075<br>Microsoft Corp. | 104.44.19.169<br>AS8075<br>Microsoft Corp. |
| ttl = 24 | * | | | | | | | | * | * | 104.44.16.180<br>AS8075<br>Microsoft Corp. | 104.44.16.214<br>AS8075<br>Microsoft Corp. |
| ttl = 25 | * | | | | | | | | * | * | 104.44.21.224<br>AS8075<br>Microsoft Corp. | 104.44.22.60<br>AS8075<br>Microsoft Corp. |
| ttl = 26 | 3.26.215.12<br>(Sydney)<br>Amazon AWS | | | | | | | | * | * | * | * |
| ttl = 27 | | | | | | | | | * | * | * | * |
| ttl = 28 | | | | | | | | | 20.5.0.129<br>(Sydney)<br>Microsoft Azure | * | * | * |
| ttl = 29 | | | | | | | | | | * | * | * |
| ttl =30 | | | | | | | | | | * | * | * |
| ttl = 31 | | | | | | | | | | 20.199.11.24<br>(Paris)<br>Microsoft Azure | * | * |
| ttl = 32 | | | | | | | | | | | * | * |
| ttl = 33 | | | | | | | | | | | * | * |
| ttl = 34 | | | | | | | | | | | * | * |
| ttl = 35 | | | | | | | | | | | * | * |
| ttl = 36 | | | | | | | | | | | * | 191.234.198.54<br>(São Paulo)<br>Microsoft Azure |
| ttl = 37 | | | | | | | | | | | * | |
| ttl = 38 | | | | | | | | | | | 20.115.40.63<br>(Virginia)<br>Microsoft Azure | |

**Table 5: Application traceroute results with `torrentdada.com` in Seoul, South Korea (*experiments conducted Dec. 2022*).**

| Hops | Traceroutes to Control Servers | | | | | | | | | | | |
|---|---|---|---|---|---|---|---|---|---|---|---|---|
| | Amazon AWS | | | | Google Cloud | | | | Microsoft Azure | | | |
| | Sydney | Paris | Virginia | São Paulo | Sydney | Paris | Virginia | São Paulo | Sydney | Paris | Virginia | São Paulo |
| ttl = 1 | 10.238.0.1 | 10.238.0.1 | 10.238.0.1 | 10.238.0.1 | 10.238.0.1 | 10.238.0.1 | 10.238.0.1 | 10.238.0.1 | 10.238.0.1 | 10.238.0.1 | 10.238.0.1 | 10.238.0.1 |
| ttl = 2 | * | * | * | * | * | * | * | * | * | * | * | * |
| ttl = 3 | 10.66.7.17 | 10.66.7.5 | 10.66.7.7 | 10.66.7.7 | 10.66.6.237 | 10.66.6.247 | 10.66.6.227 | 10.66.6.245 | 10.66.7.7 | 10.66.6.247 | 10.66.7.21 | 10.66.6.237 |
| ttl = 4 | 138.197.249.22<br>DigitalOcean | 138.197.249.18<br>DigitalOcean | 138.197.249.14<br>DigitalOcean | 138.197.249.0<br>DigitalOcean | 138.197.249.0<br>DigitalOcean | 138.197.249.18<br>DigitalOcean | 138.197.249.22<br>DigitalOcean | 138.197.249.0<br>DigitalOcean | 138.197.249.0<br>DigitalOcean | 138.197.249.22<br>DigitalOcean | 138.197.249.22<br>DigitalOcean | 138.197.249.22<br>DigitalOcean |
| ttl = 5 | 219.65.110.189<br>AS4755<br>TATA Comm. | 219.65.110.185<br>AS4755<br>TATA Comm. | 219.65.110.189<br>AS4755<br>TATA Comm. | 219.65.110.185<br>AS4755<br>TATA Comm. | 219.65.110.185<br>AS4755<br>TATA Comm. | 219.65.110.189<br>AS4755<br>TATA Comm. | 219.65.110.189<br>AS4755<br>TATA Comm. | 219.65.110.185<br>AS4755<br>TATA Comm. | 202.56.198.57<br>AS9498<br>Bharti Airtel | 202.56.198.29<br>AS9498<br>Bharti Airtel | 202.56.198.29<br>AS9498<br>Bharti Airtel | 202.56.198.29<br>AS9498<br>Bharti Airtel |
| ttl = 6 | * | * | * | * | * | * | * | * | * | 116.119.57.97<br>AS9498<br>Bharti Airtel | * | * |
| ttl = 7 | 180.87.36.9<br>AS6453<br>TATA Comm.<br>(America) | 180.87.39.25<br>AS6453<br>TATA Comm.<br>(America) | 180.87.39.25<br>AS6453<br>TATA Comm.<br>(America) | 180.87.39.25<br>AS6453<br>TATA Comm.<br>(America) | 121.240.1.46<br>AS4755<br>TATA Comm. | 121.240.1.46<br>AS4755<br>TATA Comm. | 121.240.1.46<br>AS4755<br>TATA Comm. | 121.240.1.46<br>AS4755<br>TATA Comm. | 116.119.109.205<br>AS9498<br>Bharti Airtel | * | 116.119.104.151<br>AS9498<br>Bharti Airtel | * |
| ttl = 8 | 180.87.36.41<br>AS6453<br>TATA Comm.<br>(America) | * | 180.87.39.21<br>AS6453<br>TATA Comm.<br>(America) | 180.87.39.21<br>AS6453<br>TATA Comm.<br>(America) | 34.151.125.165<br>(Sydney)<br>Google Cloud | 34.163.60.19<br>(Paris)<br>Google Cloud | 35.245.157.97<br>(Virginia)<br>Google Cloud | 35.247.224.42<br>(São Paulo)<br>Google Cloud | **Censor**:<br>116.119.94.30<br>AS9498<br>Bharti Airtel | * | **Censor**:<br>116.119.94.32<br>AS9498<br>Bharti Airtel | 182.79.239.193<br>AS9498<br>Bharti Airtel |
| ttl = 9 | 180.87.7.18<br>AS6453<br>TATA Comm.<br>(America) | 80.231.131.1<br>AS6453<br>TATA Comm.<br>(America) | 80.231.130.106<br>AS6453<br>TATA Comm.<br>(America) | * | | | | | | 198.200.130.17<br>AS8075<br>Microsoft Corp. | | **Censor**:<br>116.119.57.158<br>AS9498<br>Bharti Airtel |
| ttl = 10 | * | 80.231.62.57<br>AS6453<br>TATA Comm.<br>(America) | 66.110.96.62<br>AS6453<br>TATA Comm.<br>(America) | 66.110.96.62<br>AS6453<br>TATA Comm.<br>(America) | | | | | | **Censor**:<br>104.44.41.235<br>AS8075<br>Microsoft Corp. | | |
| ttl = 11 | * | 80.231.20.82<br>AS6453<br>TATA Comm.<br>(America) | 66.110.96.58<br>AS6453<br>TATA Comm.<br>(America) | 66.110.96.58<br>AS6453<br>TATA Comm.<br>(America) | | | | | | | | |
| ttl = 12 | * | * | * | * | | | | | | | | |
| ttl = 13 | * | * | * | * | | | | | | | | |
| ttl = 14 | * | * | * | * | | | | | | | | |
| ttl = 15 | * | * | * | * | | | | | | | | |
| ttl = 16 | * | * | * | * | | | | | | | | |
| ttl = 17 | * | * | * | * | | | | | | | | |
| ttl = 18 | * | * | * | * | | | | | | | | |
| ttl = 19 | * | * | * | * | | | | | | | | |
| ttl = 20 | * | * | * | * | | | | | | | | |
| ttl = 21 | * | * | * | * | | | | | | | | |
| ttl = 22 | * | 35.180.190.69<br>(Paris)<br>Amazon AWS | * | 54.240.244.102<br>AS16509<br>Amazon.com | | | | | | | | |
| ttl = 23 | * | | * | * | | | | | | | | |
| ttl = 24 | 3.26.215.12<br>(Sydney)<br>Amazon AWS | | * | * | | | | | | | | |
| ttl = 25 | | | 54.197.194.180<br>(Virginia)<br>Amazon AWS | * | | | | | | | | |
| ttl = 26 | | | | 18.228.203.42<br>(São Paulo)<br>Amazon AWS | | | | | | | | |

**Table 6: Application traceroute results for `cckerala.com` in Bangalore, India (*experiments conducted Dec. 2022*).**